\documentclass[twocolumn,english,twocolumn,showpacs,preprintnumbers,amsmath,amssymb,euscript,amsfonts]{revtex4}
\usepackage[T1]{fontenc}
\usepackage[latin9]{inputenc}
\usepackage{mathrsfs}
\usepackage{amsmath}
\usepackage{amssymb}
\usepackage{graphicx}
\usepackage{esint}

\makeatletter
\@ifundefined{textcolor}{}
{%
 \definecolor{BLACK}{gray}{0}
 \definecolor{WHITE}{gray}{1}
 \definecolor{RED}{rgb}{1,0,0}
 \definecolor{GREEN}{rgb}{0,1,0}
 \definecolor{BLUE}{rgb}{0,0,1}
 \definecolor{CYAN}{cmyk}{1,0,0,0}
 \definecolor{MAGENTA}{cmyk}{0,1,0,0}
 \definecolor{YELLOW}{cmyk}{0,0,1,0}
 }

\makeatother

\usepackage{babel}
\begin{document}

\title{A mechanism to pin skyrmions in chiral magnets}

\author{Ye-Hua Liu and You-Quan Li}

\affiliation{Zhejiang Institute of Modern Physics and Department of Physics, Zhejiang University,
Hangzhou 310027, People's Republic of China}
\begin{abstract}
We propose a mechanism to pin skyrmions in chiral magnets by introducing local maximum
of magnetic exchange strength, which can be realized in chiral magnetic thin films
by engineering the local density of itinerate electrons. Thus we find a way to artificially
control the position of a single skyrmion in chiral magnetic thin films. The stationary
properties and the dynamical pinning and depinning processes of an isolated skyrmion
around a pinning center are studied. We do a series of simulations to show that the
critical current to depin a skyrmion has linearly dependence on the pinning strength.
We also estimate the critical current to have order of magnitude $10^{7}\sim10^{8}\mbox{A/m}^{2}$.
\end{abstract}


\maketitle

\section{introduction}

Vortices have long been the standard topological objects in condensed matter physics.
They are topologically stable to perturbations, localized in space, and they can
be moved, created or annihilated. In recent years, another kind of topological object
{}``skyrmion'' is observed in chiral magnets. A skyrmion is a configuration of
three dimensional unit vectors in a two dimensional space which wraps around the
unit sphere once when we move from the center of the space to infinity. Skyrmions
were first proposed in high energy physics \cite{Skyrme62} and considered as an
important class of excitations in ferromagnets in condesed matter physics. Recently
they are theoretically predicted \cite{Pfleiderer06} and experimentally observed
in the ground state configuration of chiral magnets in a magnetic field \cite{Pappas09,Pfleiderer09a,Tokura10,Tokura11}.
The skyrmions either appear isolated or form a close-packed lattice. The skyrmion
lattice is also observed in a single layer of metal film \cite{Heinze11} with each
skyrmion having atomic size. Furthermore, a multiferroic skyrmion lattice in which
spin couples to electric dipole is reported very recently \cite{Tokura12}. The phase
diagram and the spinwave spectrum of skyrmion lattice in chiral magnets are very
well understood in the framework of a classical Heisenberg spin model with Dzyaloshinsky-Moriya
(DM) interaction \cite{Dzyaloshinskii58,Moriya60,Tokura10,Han10,Mochizuki12,Tchernyshyov12,Shekhtman92,Li11}. 

A skyrmion carries a topological charge, and in a conducting chiral magnet this topological
charge is a quantized emergent magnetic flux in the view of an adiabatically moving
electron. In real materials, the diameter of a skyrmion is around tens of nanometers
and the averaged magnetic field is around several Teslas \cite{Pfleiderer12}, so
the coupling between the skyrmions and the conducting electrons is quite strong.
It is observed that this emergent magnetic field gives rise to anomalous Hall effect
signal in the A-phase of chiral magnet MnSi \cite{Onose09,Pfleiderer09b}. The counteractive
of the conducting electrons to the skyrmion is the spin-transfer-torque \cite{Zhang04,Maekawa05,Zhang09,Tatara04,Pfleiderer10,Garst11,Garst12}.
It is predicted and observed that this torque will push the skyrmions to move in
the same direction of the current \cite{Zhang04,Zang11,Pfleiderer12}. It is shown
\cite{Zang11} that the kinetic term in the Lagrangian of a single skyrmion makes
the skyrmion attain a transverse motion with respect to the applied forces, and the
averaged current-skyrmion interaction gives rise to a skyrmion motion parallel to
the applied electric current. A moving skyrmion can be thought of as a moving magnetic
flux and a static magnetic field in a moving frame induces an electric field in the
laboratory frame according to Faraday's Law. This emergent electric field is measured
by Schulz \emph{et. al.} very recently \cite{Pfleiderer12}. 

Ultimately we want to manipulate these skyrmions in artificial devices. Very recenlty,
Tchoe and Han showed that a circular current can generate single skyrmions without
the aid of magnetic field \cite{Han12}. Given that a skyrmion is created, another
important point of manipulation is to put a skyrmion at the precise place we want
it to be. So the pinning and depinning effects of skyrmions in chiral magnets is
an important issue. In this paper, we propose a pinning mechanism of skyrmions by
introducing local maximum of the magnetic exchange strength. We study the effect
of such a pinning center by both stationary analysis by the energy functional and
dynamical analysis by the Landau-Lifshitz-Gilbert (LLG) equation. Then we find a
way to manipulate the skyrmions by electric current in an artificial pinning center
lattice, which may be realized by putting patterned metal grains on the surface of
a chiral magnetic thin film. We orgnize the paper as follows. In section II, we give
the field theory description of chiral magnets and the stationary analysis. In section
III, we formulate the dynamics of chiral magnets by LLG equation on a square lattice
and show the results of the simulations of the dynamical pinning and depinning processes.
In section IV, we give some discussion and further outlook.

\section{stationary analysis}

\subsection{Chiral magnet with a pinning center}

The energy density for the chiral magnet is \cite{Han10}: 
\begin{equation}
\mathscr{F}_{\mbox{tot}}\!=\!\frac{J\!\left(\boldsymbol{r}\right)}{2}\!\sum_{\mu}\!\left(\partial_{\mu}\boldsymbol{n}\right)^{2}\!+\! D\!\left(\boldsymbol{r}\right)\!\boldsymbol{n}\cdot\!\left(\nabla\!\times\!\boldsymbol{n}\right)\!-\!\boldsymbol{B}\!\cdot\!\boldsymbol{n},
\end{equation}
where $\boldsymbol{n}\left(\boldsymbol{r}\right)$ represents the orientation of
local magnetic moment with $\boldsymbol{n}^{2}=1$, $\mu=x,y,z$, $J\left(\boldsymbol{r}\right)$
refers to the local ferromagnetic exchange strength, $D\left(\boldsymbol{r}\right)$
the local strength of DM interaction and $\boldsymbol{B}$ the uniform applied magnetic
field. In previous works by mean-field theories and numerical simulations we already
know, for a certain range of $\boldsymbol{B}$, that a skyrmion solution is stable
\cite{Han10,Pfleiderer06,Li11} and the trial function for a single skyrmion is:
\begin{equation}
\boldsymbol{n}\left(\rho,\phi,z\right)=\sin\left[\theta\left(\rho\right)\right]\hat{\phi}+\cos\left[\theta\left(\rho\right)\right]\hat{z},
\end{equation}
where we have adopted cylindrical coordinate system with local unit vectors $\hat{\rho}$,
$\hat{\phi}$ and $\hat{z}$. $\theta$ is the angle between $\boldsymbol{n}$ and
$\hat{z}$.

To study the pinning effect caused by inhomogeneity, we let $J$ and $D$ be space
dependent. For simplicity, we consider an axisymmetric case such that $J$ and $D$
are of $\rho$-dependence merely. We choose 
\begin{equation}
J\left(\rho\right)=J\left(\infty\right)\left\{ 1+\lambda\exp\left[-\left(\frac{\rho}{\xi}\right)^{2}\right]\right\} \label{eq:J_form}
\end{equation}
to represent a local maximum of magnetic exchange strength. The local magnetic exchange
strength is determined by the local density of itinerate electrons \cite{Zang11}.
Here we let $D\left(\rho\right)$ and $J\left(\rho\right)$ have constant ratio $\kappa=D/\left(2J\right)$
because the density of itinerate electrons will not change the ratio of spin dependent
hopping integral to spin independent hopping integral in the underlying Hubbard model.
Introducing $\beta=B/\left[2J\left(\infty\right)\right]$, we can rewrite the energy
density as:
\begin{eqnarray}
\frac{\mathscr{F}_{\mbox{tot}}}{2J\left(\infty\right)}\! & \!=\! & \!\frac{J\left(\rho\right)}{J\left(\infty\right)}\!\left[\!\left(\!\frac{1}{2}\!\frac{\partial\theta}{\partial\rho}\!+\!\kappa\!\right)^{2}\!-\!\frac{\cos\left(2\theta\right)}{8\rho^{2}}\!+\!\frac{\kappa\sin\left(2\theta\right)}{2\rho}\!\right]\!\nonumber \\
 &  & -\beta\cos\left(\theta\right),\label{eq:energy1}
\end{eqnarray}
where we write $\theta\left(\rho\right)$ as $\theta$ in Eq. \ref{eq:energy1} and
the following Eq. \ref{eq:energy2} for compactness.

Substracting the energy density of ferromagnetic state and multiplying the measure
in space integral, we obtain the reduced one dimensional energy density $\mathscr{F}$:
\begin{eqnarray}
\frac{\mathscr{F}}{4\pi\rho}\! & \!=\! & \!\frac{J\left(\rho\right)}{J\left(\infty\right)}\!\left[\!\left(\!\frac{1}{2}\!\frac{\partial\theta}{\partial\rho}\!+\!\kappa\!\right)^{2}\!-\!\kappa^{2}\!+\!\frac{\sin\left(\theta\right)^{2}}{4\rho^{2}}\!+\!\frac{\kappa\sin\left(2\theta\right)}{2\rho}\!\right]\!\nonumber \\
 &  & -\beta\left[\cos\left(\theta\right)-1\right],\label{eq:energy2}
\end{eqnarray}
which satisfies $\int_{0}^{\infty}d\rho\mathscr{F}=\frac{1}{J\left(\infty\right)}\int d\boldsymbol{r}\left(\mathscr{F}_{\mbox{tot}}-\mathscr{F}_{\mbox{fe}}\right)$.

To get the optimized function $\theta\left(\rho\right)$ that minimizes the total
energy, we construct the {}``imaginary time evolution'': 
\begin{equation}
\frac{\partial\theta\left(\rho,\tau\right)}{\partial\tau}=-\left\{ \frac{\partial\mathscr{F}}{\partial\theta\left(\rho\right)}-\frac{\partial}{\partial\rho}\frac{\partial\mathscr{F}}{\partial\left[\partial_{\rho}\theta\left(\rho\right)\right]}\right\} ,
\end{equation}
which is a second order partial differential equation for function $\theta\left(\rho\right)$
that should be supplemented with appropriate initial condition. The boundary condition
is chosen as: 
\[
\begin{cases}
\theta\left(0,\tau\right) & =\pi\\
\theta'\left(R_{\mbox{cut}},\tau\right) & =0,
\end{cases}
\]
which means the spins point downward at the center of a skyrmion and have the same
$\theta$-angle far away from the center. The number $R_{\mbox{cut}}$ is the numerical
cut of the theoretically infinite system. The initial condition is chosen randomly
as: 
\[
\theta\left(\rho,0\right)=\begin{cases}
\pi & \rho=0\\
\mbox{random}\left(0,\pi\right) & \rho\neq0,
\end{cases}
\]
where $\mbox{random}\left(0,\pi\right)$ generates a random number from $0$ to $\pi$.
We choose $R_{\mbox{cut}}=10$ and slice the space interval $\left[0,R_{\mbox{cut}}\right]$
to $1000$ pieces, then we approximate the space derivative by finite difference
and do time integral using Runge-Kutta method. We numerically integrate this initial-boundary-value
problem until $\theta\left(\rho,\tau\right)$ does not change with time. The results
for different random initial conditions are the same, so we find the global minimum
of the energy functional and the globally optimized function $\theta\left(\rho\right)$.

\subsection{Numerical results}

Now we briefly introduce the phase diagram of a two dimensional chiral magnet in
zero temperature. For a uniform system with constant $J$ and $D$ in space, as we
increase the magnetic field from zero, the ground state changes from helical state
to skyrmion lattice state and then to field polarized ferromagnetic state \cite{Han10,Li11,Mochizuki12,Pfleiderer09a,Tokura10}.
In the phase boundary of skyrmion lattice phase and field polarized ferromagnetic
phase, an isolated skyrmion has averaged energy density that is the same as ferromagnetic
state, so statistically, a dilute population of isolated skyrmions is allowed. This
so-called skyrmion-gas state is the state of interest in the following discussion.
It is worth noticing that there is a strong resemblance between this skyrmion gas
and the vortices in the type-II superconductor at the lower critical magnetic field
$H_{\mbox{c1}}$.

\begin{figure}
\includegraphics[width=1\columnwidth]{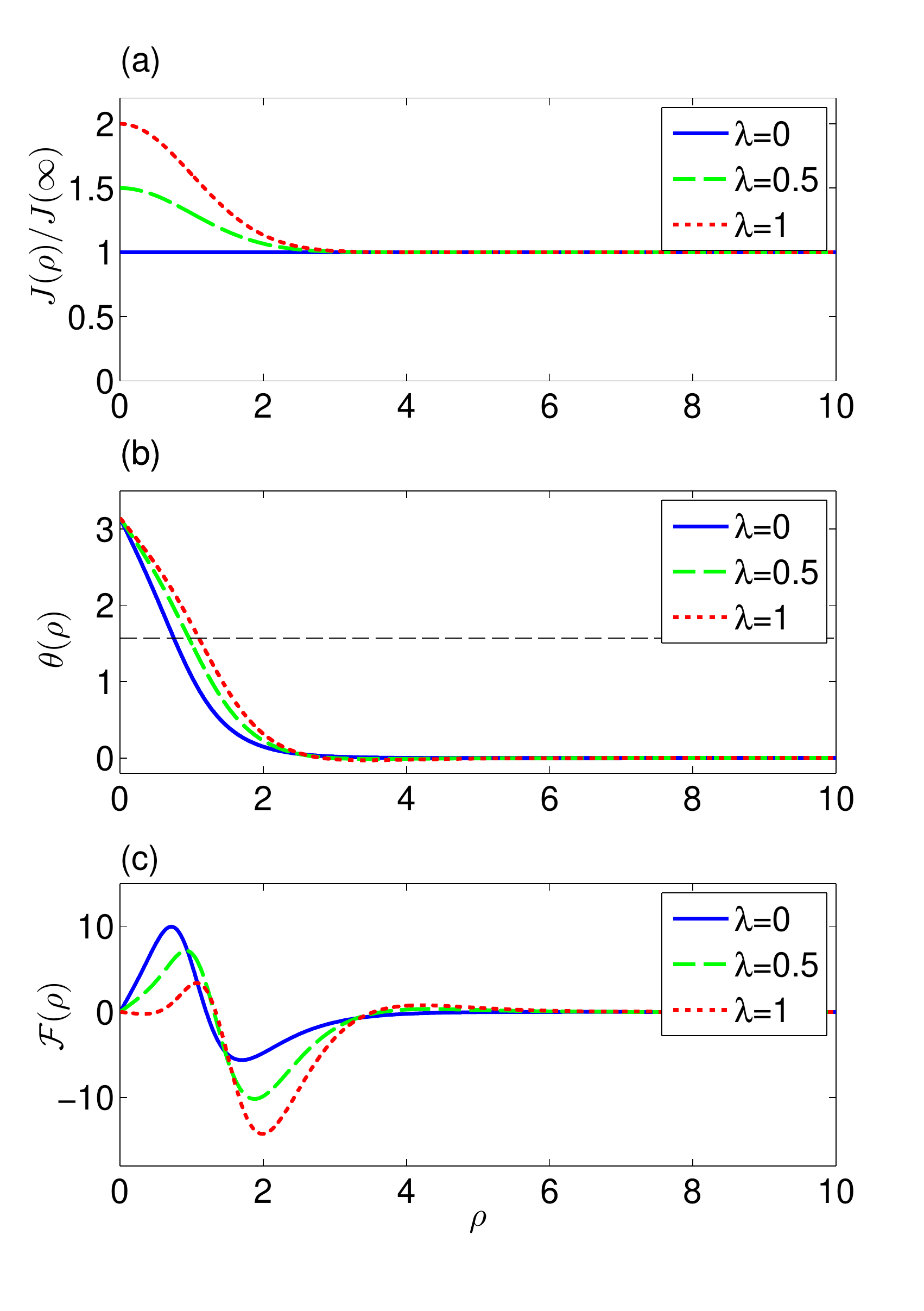}

\caption{(Color online) $\rho$-dependence of exchange strength $J\left(\rho\right)$ (a),
optimized trial function $\theta\left(\rho\right)$ (b) and energy density $\mathscr{F}\left(\rho\right)$
(c) for different pinning strength $\lambda$. It is clearly shown that a local maximum
of $J$ lowers the core energy of a skyrmion. So a skyrmion is more likely to stay
at this pinning center. The dashed line in (b) indicates $\theta=\pi/2$. \label{fig:vsrho} }
\end{figure}

\begin{figure}
\includegraphics[width=1\columnwidth]{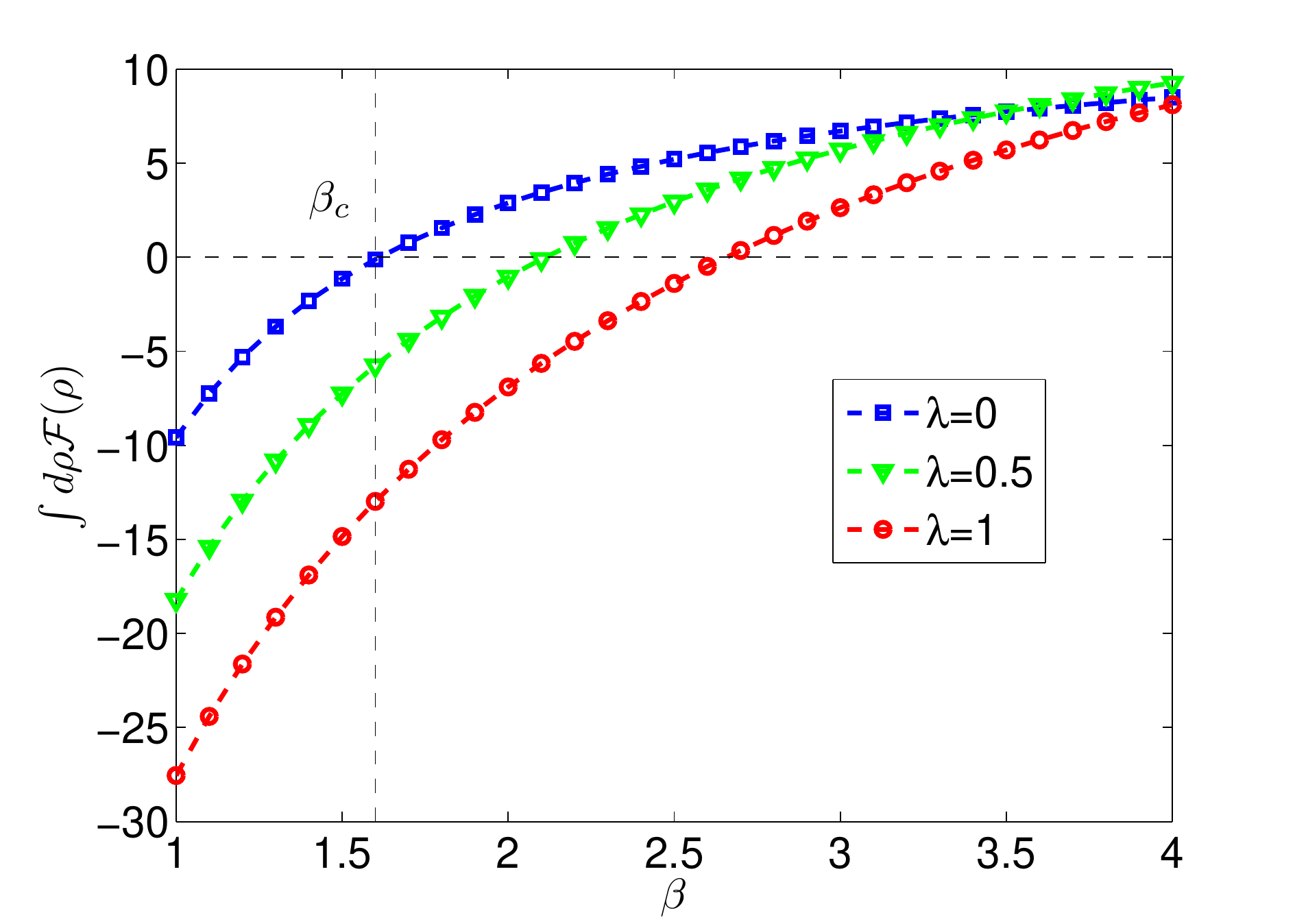}\caption{(Color online) Single skyrmion energy versus magnetic field for different strength
of pinning center $\lambda$. It is shown that a larger strength of pinning $\lambda$
gives a lower skyrmion energy $\int d\rho\,\mathscr{F}$ for the same magnetic field.
So for a pinned skyrmion, it is harder to polarize it by magnetic field. \label{fig:vsbeta} }
\end{figure}

Below we show the $\rho$-dependence of coupling $J$, optimized trial function $\theta\left(\rho\right)$
and energy density $\mathscr{F}$ in different parameters (Fig. \ref{fig:vsrho}).
We choose $\kappa=1$, $\beta=1.6$, and $\xi=2$ for three pinning strength $\lambda=0$,
$\lambda=0.5$ and $\lambda=1$. We choose $\xi=2$ so that the size of the pinning
center has the same order of magnitude as the radius of a skyrmion (so we can make
patterned pinning centers as dense as possible, see Section. \ref{sub:depin}). It
is shown that for the uniform case $\lambda=0$, the energy density profile has a
positive maximum and a negative minimum in the core region of a skyrmion. If we tune
the magnetic field $\beta$, there is a $\beta_{c}\approx1.6$ in which the positive
{}``energy cost'' and the negative {}``energy gain'' cancel each other (Fig.
\ref{fig:vsbeta}). This is precisely the aforementioned phase boundary of skyrmion
lattice phase and field polarized ferromagnetic phase.

If there is a local maximum of coupling $J$, \emph{i.e.}, $\lambda\neq0$, the core
energy is lowered. For the same $\beta$, a larger $\lambda$ makes the energy even
lower (Fig. \ref{fig:vsbeta}), thus there can be a negative energy skyrmion in the
skyrmion gas phase (even in field polarized ferromagnetic phase). This pinned skyrmion
becomes more stable than a {}``free skyrmion'' from an energy point of view. This
is what we call the pinning effect of a skyrmion by a local maximum of magnetic exchange
strength. Besides lowering the energy, the local pinning center also makes the radius
of a skyrmion a little larger.

\section{dynamical analysis}

\subsection{Landau-Lifshitz-Gilbert equation}

To study the dynamics of the pinning process, we use discrete model of chiral magnets
on a square lattice with the following Hamiltonian \cite{Han10,Han12,Li11,Mochizuki12}:
\begin{eqnarray}
H & = & \sum_{\boldsymbol{r},\vec{\delta}}\left[-J_{\boldsymbol{r},\vec{\delta}}\boldsymbol{S}_{\boldsymbol{r}}\cdot\boldsymbol{S}_{\boldsymbol{r}+\vec{\delta}}-D_{\boldsymbol{\boldsymbol{r}},\vec{\delta}}\vec{\delta}\cdot\left(\boldsymbol{S}_{\boldsymbol{r}}\times\boldsymbol{S}_{\boldsymbol{r}+\vec{\delta}}\right)\right]\nonumber \\
 &  & -\sum_{\boldsymbol{r}}\boldsymbol{B}\cdot\boldsymbol{S}_{\boldsymbol{r}},\label{eq:H for LLG}
\end{eqnarray}
where $\boldsymbol{S}_{\boldsymbol{r}}$ with $\boldsymbol{S}_{\boldsymbol{r}}^{2}=1$
are local magnetic moments at positions $\boldsymbol{r}=\left(x,y\right)$, $x,y=-\left(N-1\right)/2,...,+\left(N-1\right)/2$
for even number $N$, and $\vec{\delta}=\boldsymbol{e}_{x},\boldsymbol{e}_{y}$ are
two bond vectors in two dimensions. The dynamics of this magnetic system subjected
to an applied electric current is obtained by Landau-Lifshitz-Gilbert (LLG) equation
with a current term: 
\begin{eqnarray}
\frac{\partial\boldsymbol{S}_{\boldsymbol{r}}}{\partial t} & = & \boldsymbol{S}_{\boldsymbol{r}}\times\boldsymbol{H}_{\mbox{eff}}\left(\boldsymbol{r}\right)-\alpha\boldsymbol{S}_{\boldsymbol{r}}\times\frac{\partial\boldsymbol{S}_{\boldsymbol{r}}}{\partial t}\nonumber \\
 &  & -\sum_{\vec{\delta}}j_{\vec{\delta}}\boldsymbol{S}_{\boldsymbol{r}}\times\left(\frac{\boldsymbol{S}_{\boldsymbol{r}+\vec{\delta}}-\boldsymbol{S}_{\boldsymbol{r}-\vec{\delta}}}{2}\times\boldsymbol{S}_{\boldsymbol{r}}\right),
\end{eqnarray}
where $j_{\vec{\delta}}$ is the $\delta$-component of the applied electric current
and $\alpha$ the phenomenological Gilbert damping parameter. The effective magnetic
field $\boldsymbol{H}_{\mbox{eff}}\left(\boldsymbol{r}\right)$ is obtained by the
Hamiltonian as:

\begin{eqnarray}
\boldsymbol{H}_{\mbox{eff}}\left(\boldsymbol{r}\right) & = & -\frac{\partial H}{\partial\boldsymbol{S}_{\boldsymbol{r}}}\nonumber \\
 & = & \sum_{\vec{\delta}}\left(J_{\boldsymbol{r},\vec{\delta}}\boldsymbol{S}_{\boldsymbol{r}+\vec{\delta}}+J_{\boldsymbol{r}-\vec{\delta},\vec{\delta}}\boldsymbol{S}_{\boldsymbol{r}-\vec{\delta}}\right)\nonumber \\
 &  & +\sum_{\vec{\delta}}\left(D_{\boldsymbol{r},\vec{\delta}}\boldsymbol{S}_{\boldsymbol{r}+\vec{\delta}}\times\vec{\delta}-D_{\boldsymbol{r}-\vec{\delta},\vec{\delta}}\boldsymbol{S}_{\boldsymbol{r}-\vec{\delta}}\times\vec{\delta}\right)\nonumber \\
 &  & +\boldsymbol{B}.
\end{eqnarray}

The current term in the LLG equation $-\sum_{\vec{\delta}}j_{\vec{\delta}}\boldsymbol{S}_{\boldsymbol{r}}\times\left[\left(\boldsymbol{S}_{\boldsymbol{r}+\vec{\delta}}-\boldsymbol{S}_{\boldsymbol{r}-\vec{\delta}}\right)/2\times\boldsymbol{S}_{\boldsymbol{r}}\right]$
is the discrete version of the continuous term $-\left(\boldsymbol{j}\cdot\nabla\right)\boldsymbol{n}$
\cite{Han12,Zhang04,Zang11,Garst11,Garst12}. Here we take the cross product operation
two times to the finite difference term $\left(\boldsymbol{S}_{\boldsymbol{r}+\vec{\delta}}-\boldsymbol{S}_{\boldsymbol{r}-\vec{\delta}}\right)/2$
to remove the parallel component with respect to $\boldsymbol{S}_{\boldsymbol{r}}$,
so that the local spin module conserves during the time evoluton.

We simulate the dynamical pinning and depinning processes as follows: we use an $N$
by $N$ square lattice with periodic boundary condition, the site-dependent exchange
strength and DM interaction 
\begin{eqnarray}
J_{\boldsymbol{r},\vec{\delta}} & = & J_{0}\left\{ 1+\lambda\exp\left[-\frac{\left(x+1/2\right)^{2}+\left(y+1/2\right)^{2}}{\xi^{2}}\right]\right\} \nonumber \\
\frac{D_{\boldsymbol{\boldsymbol{r}},\vec{\delta}}}{J_{\boldsymbol{\boldsymbol{r}},\vec{\delta}}} & = & \sqrt{2}\tan\left(2\pi/d\right)
\end{eqnarray}
are chosen to have a Gaussian profile, where $J_{0}$ is the discrete version of
$J\left(\infty\right)$ in Eq. \ref{eq:J_form} and $d$ is the wavelength of the
chiral magnet on the lattice. Here we have shifted the position $\mathbf{r}=\left(x,y\right)$
in the definition of $J$ to the center of the elemental plaquette in square lattice
$\left(x+1/2,y+1/2\right)$ so that the periodic boundary condition is satisfied.
We numerically integrate the LLG equation using fourth-order Runge\textendash{}Kutta
method and follow the position of the skyrmion. We choose time-step $dt=0.001$ which
is small enough for our study and we simulate the time evolution from $t=0$ up to
$t=1000$.

\begin{figure}
\includegraphics[width=0.9\columnwidth]{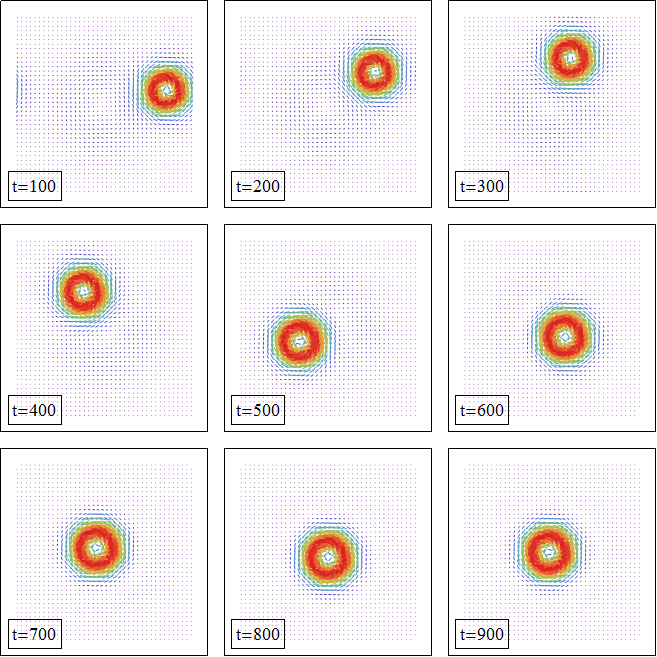}

\caption{(Color online) Snapshots of the Landau-Lifshitz-Gilbert dynamics of a skyrmion around
a pinning center on a $40\times40$ lattice. The pinning center is at the origin
point $\left(0,0\right)$. Initially we put a skyrmion at $\left(16,0\right)$, then
the offset skyrmion undergoes a circular motion around the pinning center and finally
stays at the pinning center. \label{fig:pin}}
\end{figure}

\begin{figure}
\includegraphics[width=1\columnwidth]{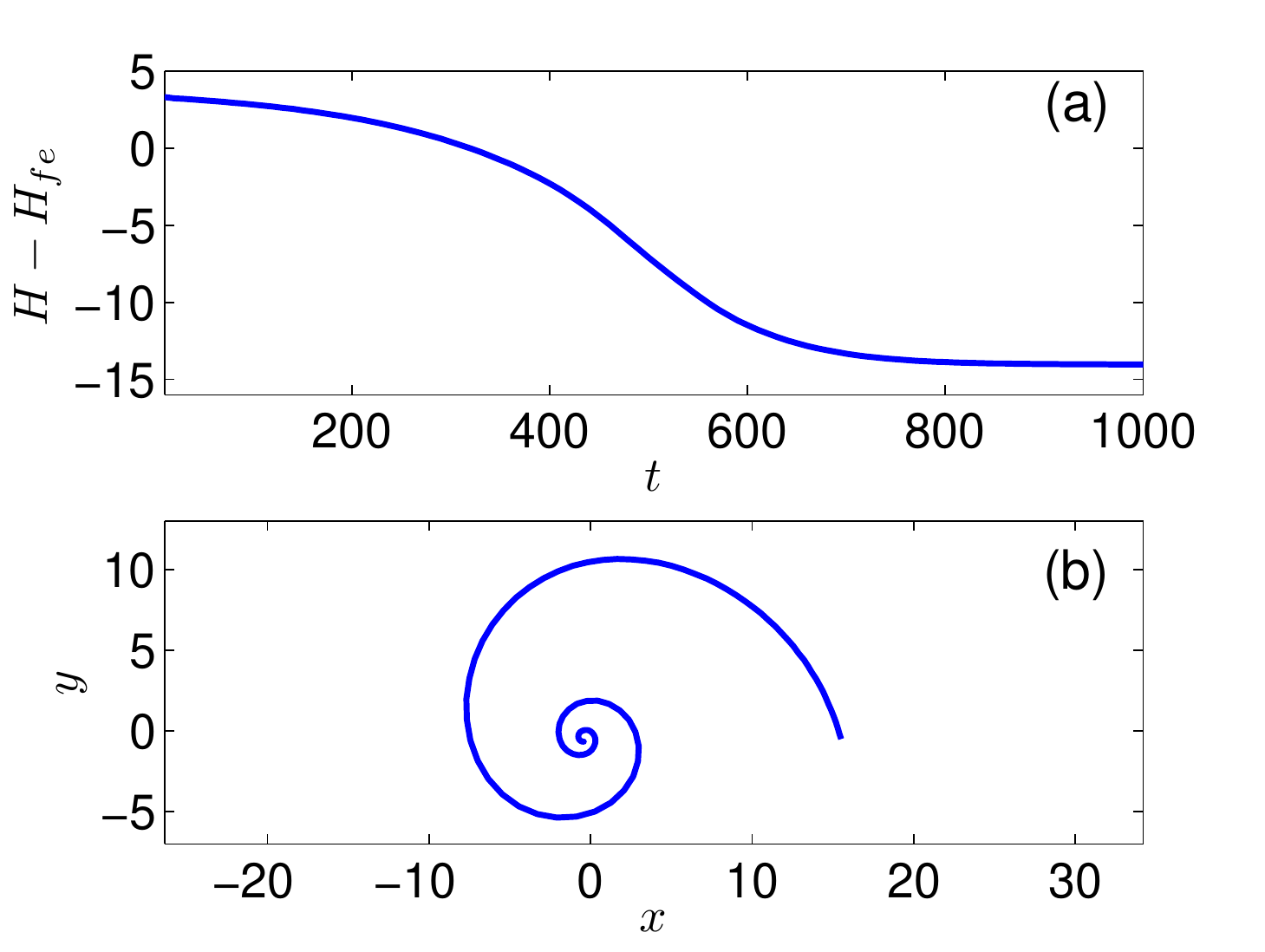}

\caption{(Color online) (a) System energy versus time for the pinning process. The energy
decreases during the whole process and finally to a stationary value. (b) The pinning
path of the skyrmion. The skyrmion moves to the pinning center circularly. \label{fig:pin-energy}}
\end{figure}

\subsection{The pinning process}

We first choose $N=40$, $J_{0}=4$, $\lambda=0.5$, $\xi=8$, $d=18$, $\boldsymbol{B}=0.9\,\boldsymbol{e}_{z}$,
and $\alpha=0.2$ with zero current. For the initial condition, we put a single skyrmion
at position $x=16$ and $y=0$ which has an offset from the pinning center $x=y=0$.
The result of the time evolution is that the skyrmion undergoes a circular motion
around the pinning center with the distance to the pinning center be smaller and
smaller and finally stays at the pinning center (Fig. \ref{fig:pin}).

We have the following heuristic explanation of this process. Because the pinned skyrmion
has lower energy than a free skyrmion and LLG equation has a damping term, the offset
skyrmion has a tendency to move to the pinning center. If we treat a skyrmion as
a point particle, the LLG dynamics of this single skyrmion has an effective kinetic
term that gives transverse motion related to the pinning force \cite{Zang11}, that
is, skyrmion motion is not parallel to the {}``attractive force'' of the pinning
center but makes an angle to the direction of the force. Combining the above two
factors we get a circular motion with smaller and smaller distance to the pinning
center.

In Fig. \ref{fig:pin-energy} we show the evolution of the system energy (minus the
energy of ferromagnetic state) versus\emph{ }time and the path of the skyrmion in
the pinning process. We can see the system energy decreases with time, this is caused
by the damping term in the LLG equation.

\begin{figure}
\includegraphics[width=0.9\columnwidth]{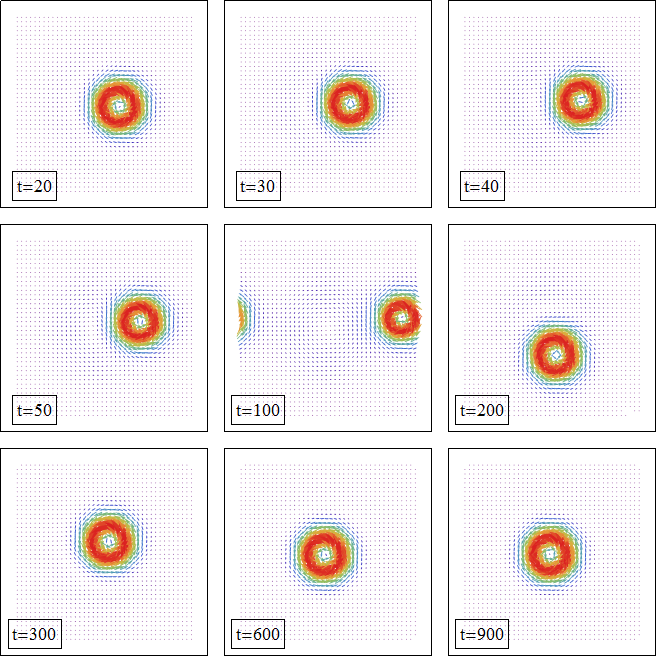}

\caption{(Color online) Snapshots of the Landau-Lifshitz-Gilbert dynamics with current for
the depinning process of a single skyrmion. We apply current $\boldsymbol{j}=\left(0.2,-0.1\right)$
to the pinned skyrmion and the skyrmion moves and leaves the unit cell from the right
boundary and enters the next unit cell from the left boundary. Then the current is
turned off (at $t_{\mbox{f}}=200$) and the skyrmion is attracted by the new pinning
center and finally is pinned. After the whole process the system state returns to
the initial one. But in fact the skyrmion is pinned by another pinning center because
we used periodic boundary condition. \label{fig:depin} }
\end{figure}

\begin{figure}
\includegraphics[width=1\columnwidth]{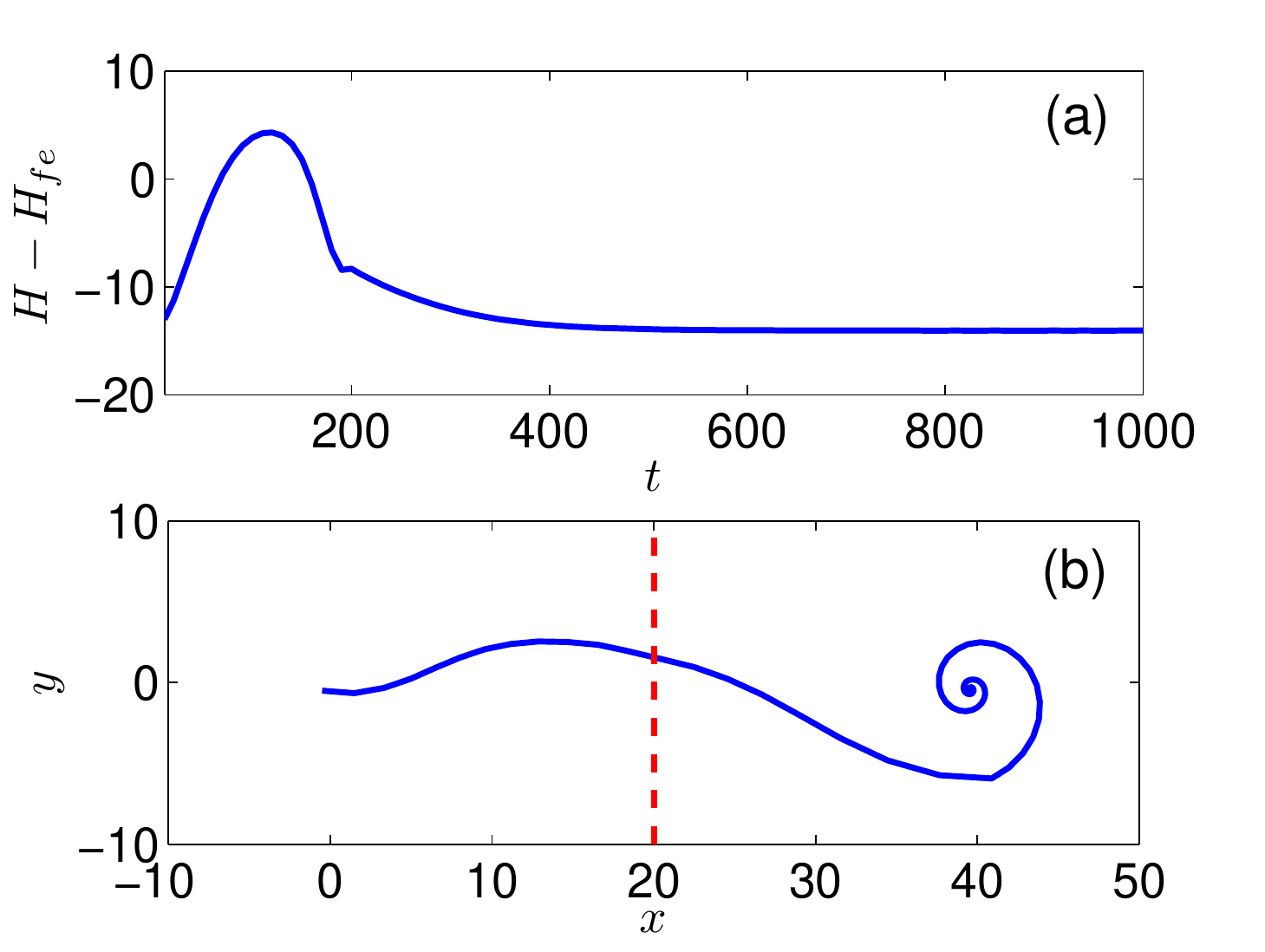}

\caption{(Color online) The evolution of the system energy versus time (a) and the depinning-pinning
path of the skyrmion (b). The current is turned on at $t=0$ and the work done by
the current enhances the system energy. After the current is turned off at $t=200$,
the system energy starts to decrease because of the damping, and finally reaches
to a stationary value which is the initial value. After the whole process the system
state returns to the initial state. In this process the skyrmion travels through
the boundary of the simulated unit cell (red dashed line). \label{fig.depin-energy}}
\end{figure}

\subsection{The depinning process \label{sub:depin}}

Now we study the depinning process of a pinned skyrmion. Because a pinned skyrmion
has lower energy than a free one, to depin a skyrmion, we need to do positive work
to it. In order to do work to a spin system, we can use a time-varying magnetic field
or an applied adiabatic electric current. The magnetic field method is often used
to sustain certain eigenmodes of the spin system near the ground state. According
to Mochizuki \cite{Mochizuki12}, a time-varying in/out-of plane magnetic field with
certain frequency can trigger the rotating/breathing internal mode of a single skyrmion.
In this paper, we choose the electric current method to push the pinned skyrmion.

To simulate the depinning process, we apply a square-current-pulse of the form 
\[
\boldsymbol{j}\left(t\right)=\boldsymbol{j}_{0}\Theta\left(t\right)\Theta\left(t_{\mbox{f}}-t\right)
\]
to the pinned skyrmion configuration, where $t_{\mbox{f}}$ is the time when we stop
the current and $\Theta\left(t\right)$ is the step-function.

We have adopted periodic boundary condition in our simulation of the LLG equation,
so we are implicitly simulating an infinite pinning center lattice in which each
pinning center lies at the origin point of our simulated $N$ by $N$ unit cell.
We expect the following phenomenon for the depinning process. When the current is
turned on, the pinned skyrmion is pushed by the current to move. The direction of
the skyrmion motion is determined by: (1) the pinning force, (2) the pushing force
(spin torque) of the current and (3) the tendency of the skyrmion to move not parallel
to the forces. Let the direction of skyrmion motion to be in the $x$-direction.
At some time the skyrmion would leave the unit cell from the right boundary then
enter the next unit cell from the left boundary and then move to the pinning center
in this unit cell. Roughly at this time we turn off the current so the skyrmion is
attracted by the pinning center in this unit cell and does circular motion just as
the above pinning process and finally stays at the new pinning center.

To simulate this process, we use the parameters $\boldsymbol{j}_{0}=\left(0.2,-0.1\right)$,
$t_{\mbox{f}}=200$ and other parameters that are the same as the pinning simulation.
We choose the current to be not parallel to the $x$-direction so that the skyrmion
moves approximately in the $x$-direction. The numerical result is in agreement with
our expectation (Fig. \ref{fig:depin}).

In Fig. \ref{fig.depin-energy}(a) we show the evolution of system energy versus
time. The system energy first increases and then decreases when the current is applied
but always decreases when the current is turned off. This is because the current
could do positive or negative work to the skyrmion depending on whether the current
is pushing the skyrmion away from or toward a pinning center. But it is the positive
work done by the current that depins the skyrmion. The system energy must decrease
without the current due to the damping. The path of skyrmion motion in the whole
process is shown in Fig. \ref{fig.depin-energy}(b), where the red dashed line indicates
the boundary of two successive unit cells of the simulated periodic lattice.

\begin{figure}
\includegraphics[width=1\columnwidth]{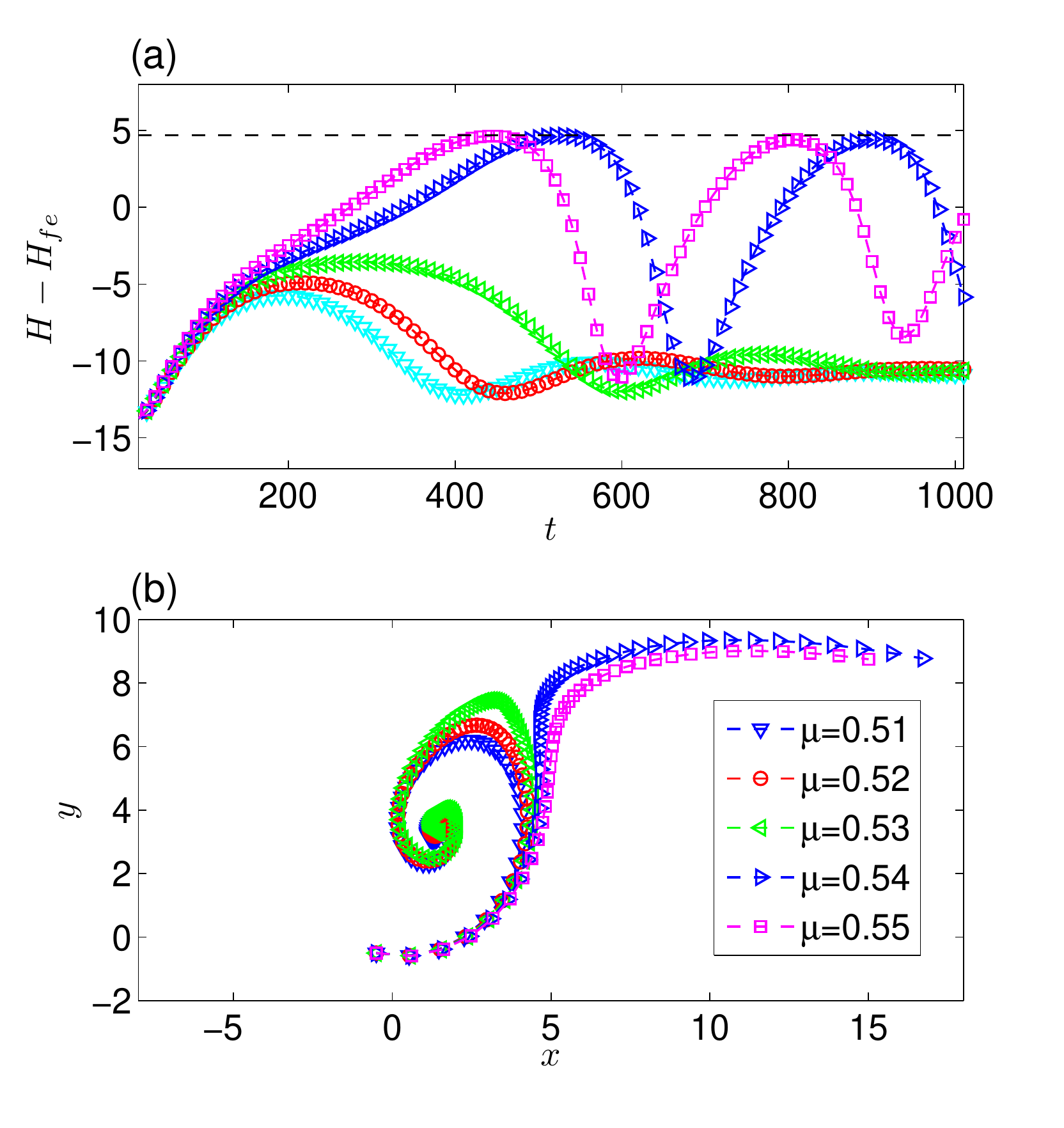}

\caption{(Color online) Energy evolution (a) and skyrmion path (b) in different applied currents.
For low currents up to $\mu=0.53$ the skyrmion cannot escape from the pinning center
at the origin point $\left(0,0\right)$ and the system energy cannot exceed the barrier
indicated by the dashed line at roughly $H-H_{\mbox{fe}}=4.7$. When the current
is as large as $\mu=0.54$, the skyrmion escapes from the pinning center and the
depinning-pinning process happens several times because we do not turn off current
in this simulation, so the system energy has a large oscillation. (parameters: $\lambda=0.5$
and $\xi=8$) \label{fig:critical current}}
\end{figure}

\begin{figure}
\includegraphics[width=1\columnwidth]{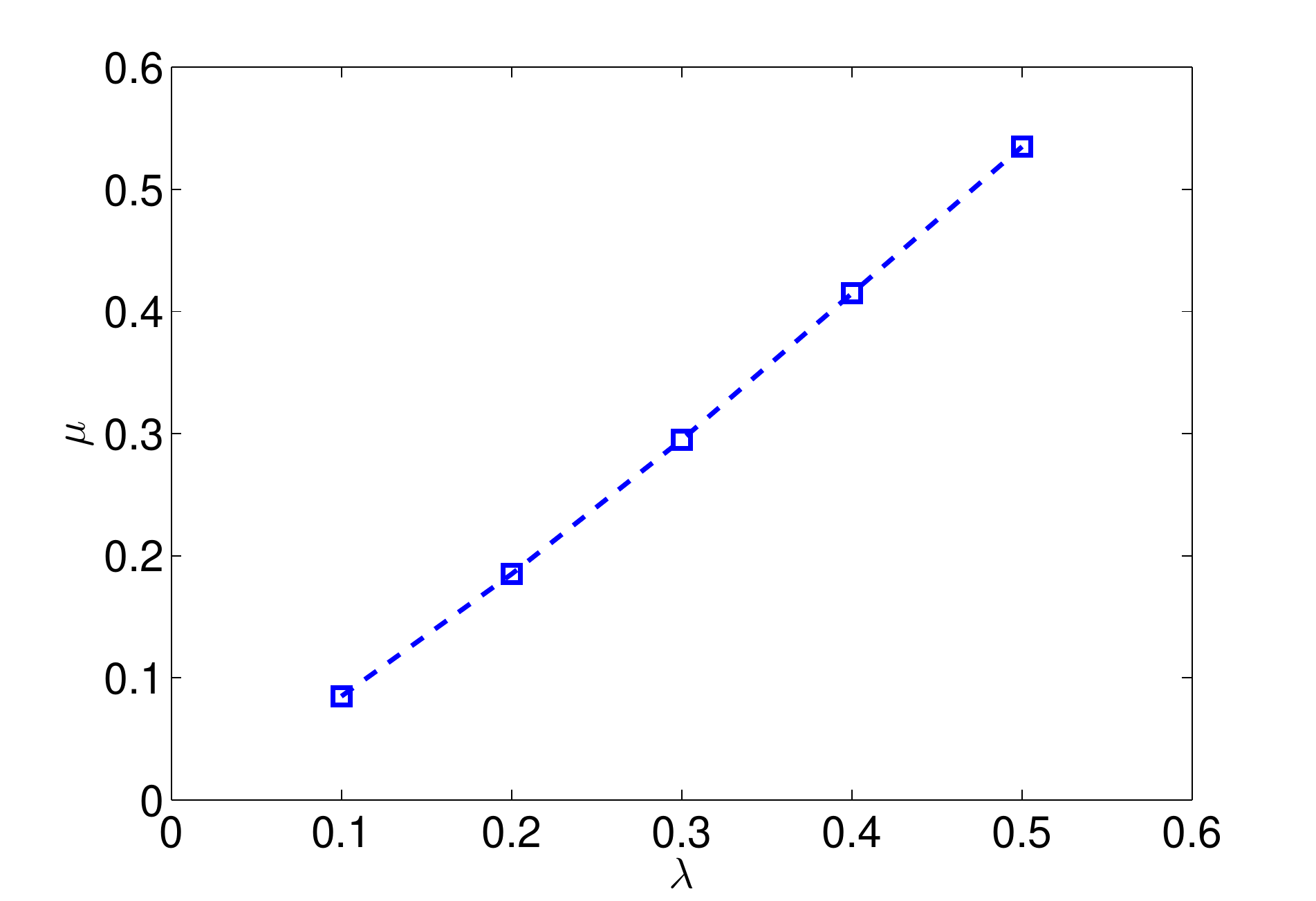}\caption{(Color online) Critical current versus pinning strength. The critical current $\boldsymbol{j}_{c}=\mu\left(0.2,-0.1\right)$
needed to depin a skyrmion varies almost linearly with respect to the pinning strength
$\lambda$. \label{fig:jc_vs_lam}}
\end{figure}

\subsection{Critical current of the depinning process}

We found in our simulation that the pinned skyrmion would not travel to other pinning
centers but instead circulates around its own pinning position and finally stays
at a nearby position if the applied current is not large enough. So there is a threshold
critical current to depin a skyrmion.

To study the critical current of depinning, we do a series of simulations with increasing
current intensity $\boldsymbol{j}=\mu\left(0.2,-0.1\right)$ with $\mu=0.51,...,0.55$
for a pinning strength $\lambda=0.5$. The result shows that there exists a critical
current below which the skyrmion cannnot escape from the pinning center but circulates
around the pinning center and the energy will not exceed the barrier energy set by
the pinning center; but when the current is larger than the critical current, the
skyrmion could escape from the pinning center and the energy exceeds the barrier
energy (Fig. \ref{fig:critical current}).

Next we simulate the critical current in different pinning stength $\lambda$ and
we find an almost linear dependence of the critical current $\boldsymbol{j}_{c}=\mu\left(0.2,-0.1\right)$
with respect to the pinning strength (Fig. \ref{fig:jc_vs_lam}). The approximate
relation is $\mu\approx\lambda$.

\subsection{Estimation of the critical current}

To estimate the critical current, we rescale the LLG in real material to our dimensionless
form. Let the letters with a prime represent physical parameters in real material.
First we do space rescaling, supposing the real material has magnetic exchange $J'$
and crystal lattice spacing $a'$, then because the magnetic exchange scales as the
inverse square of length, we have $a^{2}J=a'^{2}J'$. Here $J$ is the rescaled magnetic
exchange for our simulated space unit defined by $18a\approx\lambda'$, where $\lambda'$
is the helical wavelength in real material. Next we do time rescaling, in our simulation
we have implicitly chosen the time unit to be $t_{0}=4\hbar/J$. After these rescalings,
the unit of the electric current is $j_{0}=\left(S/p\right)\left[q_{\mbox{e}}/\left(a^{2}t_{0}\right)\right]$
where $S$ is the spin quantum number of the local magnetic moment, $p$ is the fraction
of spin polarization of the electric current. Next we substitute real parameters
chosen as $a'=0.4\mbox{nm}$, $J'\approx3\mbox{meV}$, $\lambda'\approx60\mbox{nm}$,
$p\approx0.1$ and $S\approx1$ \cite{Han12}, then we get $j_{0}\approx2.4\times10^{9}\mbox{A/m}^{2}$.
With the above result we have $j_{c}\approx5.3\lambda\times10^{8}\mbox{A/m}^{2}$
(here $\lambda$ refers to the pinning strength with order of magnitude $10^{-1}$
in our simulation) and the typical pinning and depinning time around $1000t_{0}\approx10\mbox{ns}$.

\subsection{Multiple-skyrmion depinning}

\begin{figure}
\includegraphics[width=0.9\columnwidth]{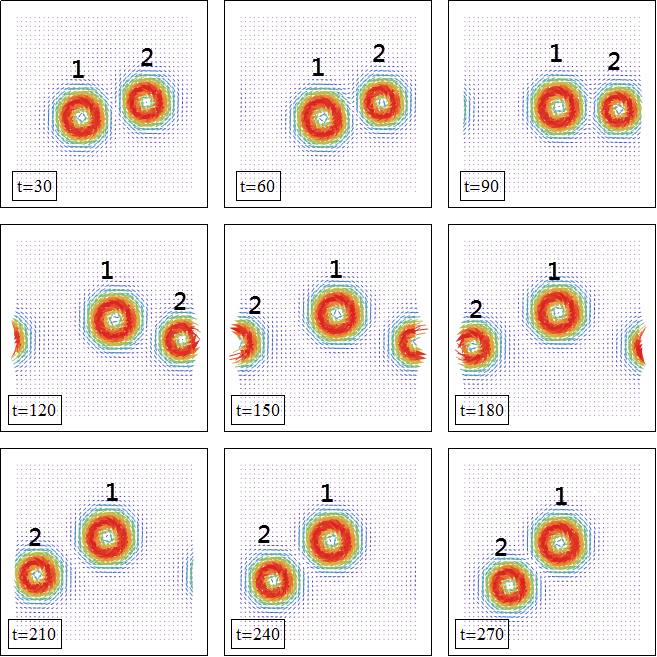}\caption{(Color online) Snapshots of two skyrmion depinning. The current is chosen to be smaller
than the critical current. But one of the two skyrmions is depinned. This is because
the depinned skyrmion is subjected to both the pinning force and the repulsive force
of the other skyrmion and the net result is the weakening of the pinning effect.
This figure clearly shows that the depinned skyrmion is smaller than the pinned one,
which is a conclusion of our stationary analysis. \label{fig:two sky depin}}
\end{figure}

If there are more than one skyrmions around a pinning center, each skyrmion will
feel simultaneously the attractive force of the pinning center and the repulsive
force of other skyrmions because skyrmions repel each other. The net result is that
the pinning effect is weakened. We simulate a two skyrmion case with $\lambda=0.5$,
$\boldsymbol{j}=0.4\left(0.2,-0.1\right)$ which is below the critical current for
a single skyrmion. The result is that one of the skyrmions is depinned by this current
(Fig. \ref{fig:two sky depin}). Generalizing to more skyrmion case, we may say that
multiple skyrmion configuration is more easily depinned. This is a well-known experimental
result \cite{Pfleiderer12}.

\subsection{Influence of the non-adiabatic spin-torque term}

Besides the current term in our LLG equation, in conducting chiral magnets, there
is another current related term which is the {}``non-adiabatic spin-torque term''
\cite{Han12,Garst11,Garst12}. Combining this term to the existing current term we
have:
\begin{equation}
\left(\boldsymbol{j}\cdot\nabla\right)\boldsymbol{S}\rightarrow\left(\boldsymbol{j}\cdot\nabla\right)\boldsymbol{S}+\beta\boldsymbol{S}\times\left[\left(\boldsymbol{j}\cdot\nabla\right)\boldsymbol{S}\right],
\end{equation}
where $\beta$ represents the strength of the non-adiabatic torque. It is known that
this term alters the trajectory of single skyrmions. When there is no pinning center,
the skyrmion would move parallel to the applied current if the value $\beta$ equals
to the Gilbert damping $\alpha$, otherwise the skyrmion motion would have a transverse
component to the right/left of the current direction if $\beta$ is larger/smaller
than $\alpha$ \cite{Han12}.

\begin{figure}
\includegraphics[width=1\columnwidth]{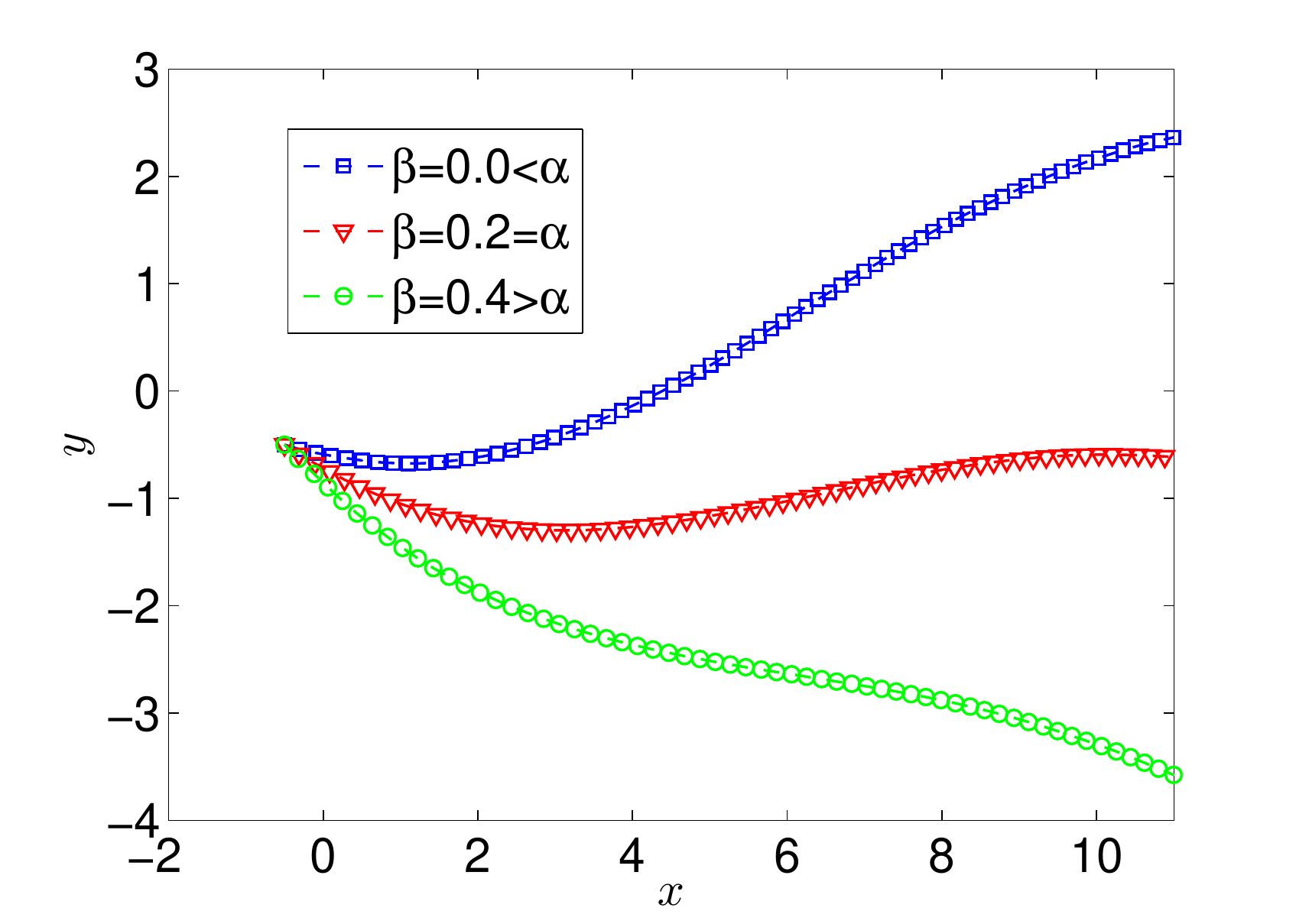}\caption{(Color online) Skyrmion trajectory for the depinning process in different \textbf{$\beta$}-values.
It shows that the main influence of the $\beta$-term is the direction of the transverse
component of motion with respect to the current direction. Because there exists a
pinning center in our simulation, even if $\beta=\alpha$, the skyrmion still moves
in a curved path. \label{fig:beta-term}}
\end{figure}

When there is a pinning center, the $\beta$-term has similar effect. In the following
simulations we still use the current $\boldsymbol{j}=\left(0.2,-0.1\right)$. In
Fig. \ref{fig:beta-term} we show that the trajectory of skyrmion in the depinning
process is altered by the $\beta$-term. Because there is a pinning center, even
if $\beta=\alpha$ the skyrmion still moves in a curved path.

\section{summary and outlook}

We proposed a mechanism to pin skyrmions by a local maximum of magnetic exchange
strength $J$ in the conventional Heisenberg model with DM interaction. Physically,
this local maximum of exchange strength is caused by a local maximum of the density
of itinerate electrons in the material. We studied the effect of this local maximum
by stationary analysis of the energy functional and found that this pinning center
makes an isolated skyrmion more stable by lowering the core energy of it. Besides
the stationary analysis, we also studied the dynamics of the chiral magnet with pinning
center by LLG equation with a current term. We found that an offset skyrmion near
the pinning center is attracted to the pinning center and undergoes a circular motion
around the pinning center and finally stays at there. This pinned skyrmion can further
be depinned by an applied current pulse, which pushes the skyrmion away from the
pinning center and let the skyrmion be attracted by other pinning centers. Our simulation
shows that there exists a critical current below which the pinned skyrmion cannot
be depinned. We did a series of simulations and found that the critical current depends
approximately linearly to the pinning strength and we estimated the critical current
to have order of magnitude $10^{7}\thicksim10^{8}\mbox{A/m}^{2}$. With the pinning
and depinning processes, we actually found a way to manipulate the skyrmions in artificially
made pinning center lattice. This pinning center lattice is expected to be realized
by putting metal grains on the surface of a chiral magnetic thin film in a certain
way.

Our pinning and depinning mechanism resembles in spirit to the mechanism of the {}``magnetic
domain-wall racetrack memory'' proposed by Parkin \emph{et. al.} \cite{Parkin08}
in which they use magnetic domain walls to represent information but here we use
single skyrmions. The main components of a racetrack memory are: (1) a writing unit,
(2) a racetrack and (3) a reading unit. For skyrmions, the writing unit could be
realized by the method of {}``skyrmion generation by a circular current'' \cite{Han12}.
Our work can be thought of as a model of the two dimensional racetrack for single
skyrmions. In Parkin's work they use patterned notches along the edges of the racetrack
to make pinning centers for the domain walls and they use electric current to depin
the domain walls to make the information {}``flow''. In our present proposal we
use patterned local maximums of exchange strength to pin skyrmions and still electric
current to depin them. The critical current to depin skyrmions from disorder is roughly
$10^{6}\mbox{A/m}^{2}$ \cite{Pfleiderer12}, which is much smaller than our estimated
critical current to depin skyrmions from artificially made pinning center $10^{8}\mbox{A/m}^{2}$.
The above two values are both much smaller than the critical current in racetrack
memory which is of order $10^{12}\mbox{A}/\mbox{m}^{2}$. Skyrmions in conducting
chiral magnets can be more easily depinned and causes much smaller Joule-heating
for potential application.\\

\begin{acknowledgments}
This work is supported by NSFC (Grant No. 11074216). The authors thank Prof. Jung
Hoon Han for careful reading of the initial manuscript and fruitful suggestions.\end{acknowledgments}

\end{document}